\documentclass[twocolumn,showpacs,prb,aps,amsmath,amssymb]{revtex4-1}
\usepackage{graphicx}

\begin{document}
\title{A critical analysis of vacancy-induced magnetism in mono and bilayer graphene}

\author{J. J. Palacios and F. Yndur\'ain}
\affiliation{Departamento de F\'{\i}sica de la Materia Condensada, Universidad Aut\'onoma de Madrid, 
Cantoblanco, Madrid 28049, Spain}

\begin{abstract}

The observation of intrinsic magnetic order in graphene and graphene-based materials relies on the formation of 
magnetic moments and a sufficiently strong mutual interaction. 
Vacancies are arguably considered the primary source of magnetic moments. Here we present an in-depth density 
functional theory study of the spin-resolved electronic structure of (monoatomic)
vacancies in graphene and bilayer graphene. We use two different methodologies: 
supercell calculations with the SIESTA code and cluster-embedded calculations with the ALACANT package.
Our results are conclusive: The vacancy-induced extended $\pi$ magnetic moments, 
which present long-range interactions and are capable of magnetic ordering, vanish at any experimentally relevant
vacancy concentration.
This holds for $\sigma$-bond passivated and un-passivated reconstructed vacancies, although, for the un-passivated ones,
the disappearance of the $\pi$ magnetic moments is accompanied by a very large
magnetic susceptibility. Only for the unlikely case of a full $\sigma$-bond passivation, 
preventing the reconstruction of the vacancy, a full value of 1$\mu_B$ for the $\pi$ extended magnetic moment is recovered for both mono and bilayer cases.  Our results put on hold claims of vacancy-induced ferromagnetic or antiferromagnetic order
in graphene-based systems, while still leaving the door open to $\sigma$-type paramagnetism. 

\end{abstract}


\maketitle

\section{Introduction}
According to theory, the existence of intrinsic (without invoking foreign species) 
magnetism in graphene and graphene-based materials should be the rule rather than the exception.  
Aside trivial paramagnetism associated with $\sigma$ dangling bonds of undercoordinated C atoms, 
a more interesting $\pi$ magnetism should arise at
zigzag edges\cite{Fujita_1996,PhysRevB.54.17954,Wakabayashi99,Hikihara03,PhysRevB.67.092406,Son06,fernandez-rossier:177204,Jiang07-1,PhysRevB.77.035411,PhysRevB.76.245415},
bulk defects such as vacancies\cite{El-Barbary03,PhysRevLett.93.187202,vozmediano:155121,yazyev:125408,palacios:195428,0953-8984-21-19-196002,Li10,Pisani08,Dharma-warda08}, or grain boundaries\cite{Cervenka:nature}, which either appear naturally or can be created in a more or 
less controlled manner. However, undisputed experimental evidence of magnetic order remains elusive since early claims of 
observation of ferromagnetism in 
irradiated graphite\cite{PhysRevB.66.024429,barzola-quiquia:161403a} and graphene \cite{Wang09}. Recent claims based on 
transport\cite{PhysRevB.83.121401} and local probe measurements\cite{PhysRevLett.104.096804,Tao11,Cervenka:nature}
seem more solid, but not entirely free from controversy\cite{PhysRevLett.105.257203}. The reasons why the observation of 
ferromagnetism in graphene and graphene derivatives is so elusive, even at low temperatures,
are still unclear, but can generically be traced back to two facts: 1) the magnetic instability leading to
the appearance of magnetic moments can be superseded by structural (Jahn-Teller) instabilities or unwanted passivation 
by foreign species, and 2) the underlying antiferromagnetic correlations inherent to graphene favor this type of magnetic order over ferromagnetism even if the magnetic moments truly exist. 

Graphene represents the paradigm of bipartite lattices. At the heart of the bipartite nature lies the reason why
some graphene derivatives result in half-filled $\pi$ states at the Fermi energy which spin-split due to 
electron-electron interactions. When these interactions are restricted to be local, as described, e.g., by a one-orbital Hubbard 
model, and the electron-hole symmetry is exactly preserved, the existence of a magnetic ground state with total 
spin $S=|N_A-N_B|/2$ is guaranteed by  a theorem by Lieb\cite{PhysRevLett.62.1201}, where $N_A-N_B$ is the difference between 
the number of atoms in each sublattice, i.e., the sublattice imbalance. This imbalance appears whenever 
graphene is cut or grown into triangular shapes bounded by zigzag edges\cite{fernandez-rossier:177204}
or, conversely, when C atoms are removed from bulk graphene, creating 
vacancies\cite{vozmediano:155121} or voids\cite{palacios:195428} with similar triangular
shapes. Interestingly, even when $N_A-N_B=0$, as is 
the case in zigzag nanoribbons\cite{Fujita_1996,PhysRevB.54.17954,Wakabayashi99,Hikihara03,PhysRevB.67.092406,Son06}, 
large hexagonal graphene nanoflakes with zigzag edges\cite{fernandez-rossier:177204,Jiang07-1}, or
voids of similar shape in bulk graphene\cite{palacios:195428}, 
magnetic solutions may appear, but always with an envelope antiferromagnetic order on top of a local ferromagnetic order. 
Smaller structures with $N_A-N_B=0$ are not magnetic as, e.g., recent work on divacancies in graphene\cite{Divacancy} shows.

\begin{figure}
\includegraphics[width=0.8\linewidth]{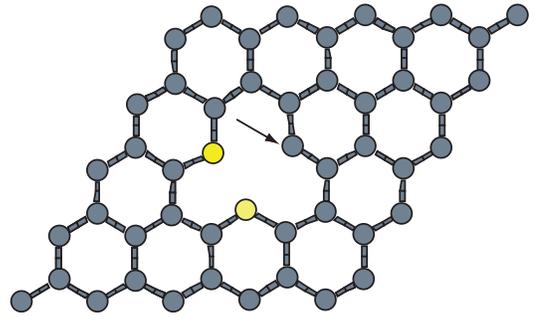}
  \caption{(Color online) Atomic structure around a single atom vacancy in graphene. The arrow indicates the atom with the dangling $\sigma$ bond and the colored atoms are those rebonded.}
\label{Atomicstructure}
\end{figure}

In between Lieb's theorem and the observation of magnetic fingerprints in graphene-based systems stands a number of 
assumptions easily overlooked, 
namely; i) that the bipartite atomic structure is preserved, ii) that hydrogenation or passivation of the
extended $\pi$ states does not occur, iii) that the Hubbard model
is a good approximation to describe graphene $\pi$ interactions, and (iv) that the substrate does not play a significant role. 
Despite that both chemical synthesis and physical approaches 
are making progress into the creation of locally or globally
imbalanced graphene structures, it still remains a remarkably challenging task.  The second condition is a major 
experimental challenge since it is difficult to avoid full passivation of the edges\cite{Wassmann-08}
(not to mention to achieve the often implicit selective passivation of the $\sigma$ bonds, 
while avoiding that of the $\pi$ states), at least under 
conditions compatible with standard magnetic measurements. Moreover, if passivation is completely 
avoided in ultra-high vacuum conditions, the equilibrium atomic structure may develop a reconstruction that ruins the first
condition and that, energy wise, competes favorably with the magnetic 
instability\cite{koskinen:115502,PhysRevB.83.045414}. Substrates may 
prevent this reconstruction from taking place\cite{tian2011}, 
but it is unclear whether or not they always respect the magnetic instability.
Finally, while the (mean-field) Hubbard model has shown its reliability in reproducing results 
obtained with more sophisticated approximations [typically density functional theory (DFT)\cite{fernandez-rossier:177204}], 
the comparison has only been
carried out in the most favorable situation, namely, saturating the $\sigma$ bonds with H, thus avoiding unwanted
lattice reconstructions. The extent to which the unsaturated
$\sigma$ bonds or lattice reconstructions may invalidate the use of the one-orbital Hubbard model remains largely unexplored.

\begin{figure}
\includegraphics[width=1.0\linewidth]{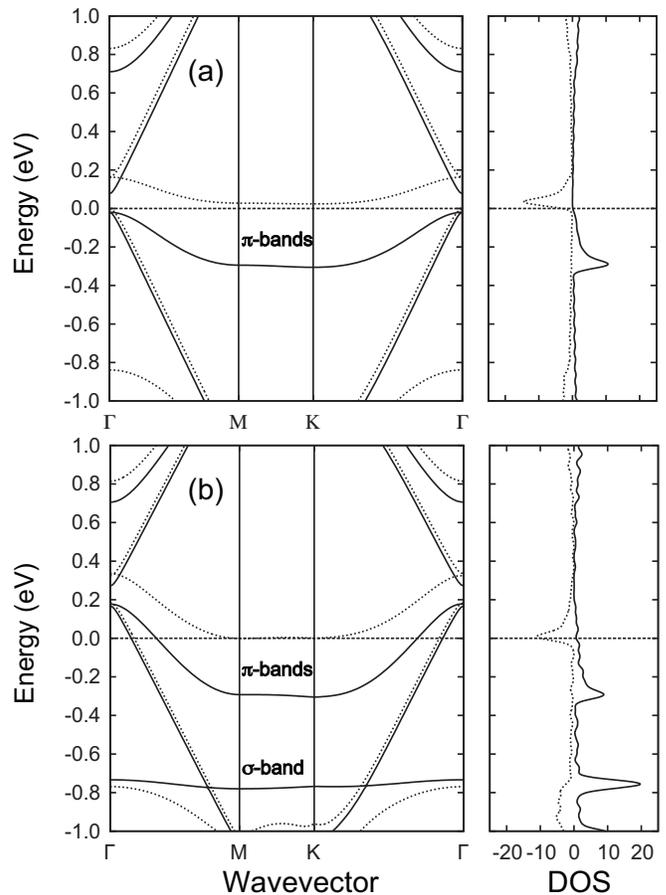}
  \caption{Calculated spin resolved band structure (left panels) and densities of states (right panels) at a vacancy
 in a 6x6 super-cell in monolayer graphene. In (a) the three $\sigma$ dangling bonds left by the removal of C 
atom are saturated with H atoms. In (b) a bare vacancy is considered.  Full relaxation of the atomic geometry is performed in both cases. Solid and broken lines indicate spin majority and spin minority
 electronic states, respectively. The zero of energy is at the Fermi level. }
\label{mono3H}
\end{figure}

While the magnetic instability at zigazg edges is under present theoretical and experimental scrutinity, 
the natural hosts of magnetism 
in bulk, vacancies, have not received due critical attention so far.
Single-atom vacancies in bulk graphene are the simplest structures complying, in principle, with the conditions for the 
appearance of extended $\pi$ magnetic moments, both in mono\cite{PhysRevLett.93.187202,vozmediano:155121,yazyev:125408,palacios:195428,Kumazaki07,0953-8984-21-19-196002,Pisani08,Dharma-warda08} and 
multilayer graphene\cite{0953-8984-20-23-235220,PhysRevLett.104.036802,faccio08}.  When H saturation
of the dangling bonds (left upon removal of a C atom)
prevents the Jahn-Teller reconstruction of the vacancy, the value of the induced magnetic moment is expected to be
 1 $\mu_B$ for the $\pi$ electrons plus 0 or 1 $\mu_B$
for the $\sigma$ bonds, depending on whether 3 or 2 H atoms are available for saturation, respectively. 
Discrepancies, however, can be found in the literature regarding the actual value of the magnetic moment when a single H (or
no H at all) saturates a dangling bond and the vacancy reconstructs 
(see Fig. \ref{Atomicstructure}).  Values for the $\pi$ magnetic moment 
ranging from $\approx 0.0\;\mu_B$ to $\approx 1.0 \; \mu_B$ have been reported 
in this case\cite{El-Barbary03,PhysRevLett.93.187202,yazyev:125408,0953-8984-21-19-196002,Dharma-warda08,2011arXiv1105.1129N}. 

To illustrate the source of the discrepancy we show in Fig. \ref{mono3H} the electronic structure of a single vacancy in a 
graphene monolayer in two different instances. In panel (a) we consider the three $\sigma$ dangling bonds,
left by the removal of a C atom, saturated with H atoms, whereas in panel (b) we consider no H passivation. In both cases 
a full atomic relaxation is carried out. A schematics of the atomic structure in the 
second case is shown in Fig. \ref{Atomicstructure} which is similar to the ones previously reported in the 
literature\cite{El-Barbary03,PhysRevLett.93.187202,yazyev:125408,0953-8984-21-19-196002,Dharma-warda08,2011arXiv1105.1129N}. 
The differences in the electronic structure between both situations 
are remarkable.  In the former case the trigonal atomic symmetry is maintained and, despite small deviations from the perfect
atomic lattice, Lieb's theorem is still expected to apply. We obtain two 
well separated spin minority and spin majority $\pi$ bands (dispersive due to the 
supercell periodicity) near the Fermi level situated at the Dirac point.
The magnetic moment associated with the $\pi$ orbitals is actually 1  $\mu_B$.
In Fig.  \ref{mono3H}(b), aside the appearance of a $\sigma$ band at -0.75 eV, the situation is different. The   $\pi$ bands 
overlap in energy close to the Fermi level and, therefore, the magnetic moment is smaller than 2  $\mu_B$ (1.71 in this case). 
Importantly, the Dirac point lies above the Fermi level at around 0.25 eV.

These results already show the subtle effect of breaking the bipartite character of the lattice by the relaxation 
of the atoms around the vacancy which, in turns, induces the breaking of the electron-hole symmetry.  
In the case of bilayer graphene the situation is expected to be even more subtle since,
along with these details, the influence of the bottom layer needs to be taken 
into consideration. Graphene layers are coupled via van der Waals interactions and, to our knowledge, the 
existing calculations did not tackle this issue appropriately. Also,  as discussed below, 
the value of the $\pi$ magnetic moment turns out to be quite sensitive to details of the calculation 
such as size and shape of the supercell, generally decreasing with size.
To compound things even further, a recent scanning tunneling spectroscopy (STS) observation of the density of states 
associated with vacancies created under controlled conditions on the surface layer of graphite  
has been 
interpreted as a manifestation of $\pi$ magnetism\cite{PhysRevLett.104.096804}. 
A second reading of the experimental data, however, can also be interpreted otherwise, as the evidence
of the {\em absence} of magnetism.  This has prompted us to question whether vacancy-induced $\pi$ magnetism exists at all.

Here we present extensive and detailed DFT calculations for vacancies in both mono and bilayer
graphene, including van der Waals interaction\cite{vdWaals1} as implemented in the SIESTA code\cite{Siesta1,Siesta2,vdWaals2}.
Our results indicate that 
reconstructed vacancies in mono and bilayer graphene can host highly localized $\sigma$ magnetic moments of 1 $\mu_B$ (i.e.,
one unpaired spin), but that {\em the overall 
extended $\pi$ magnetism progressively decreases as the size of the supercell increases, 
extrapolating to zero in the zero-concentration limit}. This limit is reached more rapidly when
the dangling $\sigma$ bond is saturated with H (and the $\sigma$ magnetic moment is quenched)
and also when the vacancy is created on the bilayer. In the latter no significant differences are appreciated between the
two lattice sites in this regard.
Calculations for a single vacancy performed with a cluster-embedded methodology as implemented in the ANT.G code\cite{ALACANT}
also yield values of the magnetic moment in the $\pi$ orbitals approaching zero, strengthening our conclusion.
On the other hand, as expected, a value of 1 $\mu_B$ for the $\pi$ magnetic moment 
is obtained for the unlikely case of a total
hydrogenation of the $\sigma$ orbitals which prevents the Jahn-Teller reconstruction and recovers the Lieb's scenario. 

\section{methodology}
Most of the calculations reported here have been performed with the SIESTA code  which uses a basis of numerical 
atomic orbitals~\cite{san} and separable~\cite{kle} norm conserving pseudopotentials~\cite{tro} with partial core 
corrections~\cite{lou}. We have found satisfactory the standard double-$\zeta$ basis with polarization orbitals (DZP) 
which has been used throughout this work. Also a ghost atom at the vacancy has been included to improve the basis set,
 although it does not change the DZP results. The convergence of the relevant precision parameters was carefully checked. 
The real space integration grid had a cut-off of 500 Ryd. Of the order of up to 600 $k$ points were used in the 
two-dimensional Brillouin zone sampling using the Monkhorst-Pack k-points sampling. Spin resolved calculations are performed in most cases.
To accelerate the self-consistency convergence, a polynomial broadening of the energy levels was performed using the 
method of Methfessel and Paxton \cite{met} which is very suitable for systems with a large variation of the density 
of states in the vicinity of the Fermi level as is the case in our system (see below). Broadening like Fermi-Dirac 
can be inappropriate and give wrong results.  It is worth mentioning that the energy differences between 
non-magnetic and magnetic solutions are, in general, small, what requires a very high convergence in all 
precision parameters and tolerances. To obtain the equilibrium geometry we relaxed all the atoms until the forces acting 
on them were smaller than 0.01 eV/\rm{\AA}. We obtain for the defect free graphene layer a nearest-neighbor distance 
of 1.435   \rm{\AA} as compared with the experimental value of 1.42  \rm{\AA}. In the bilayer calculation 
including van der Waals forces the distance between planes is 3.42  \rm{\AA} whereas the experimental one for 
graphite is 3.35  \rm{\AA}. To calculate the geometrical and electronic structure of defects, we use the supercell 
calculation method with $n\times m$ cells containing the defect for $n$ and $m$ integers and standard unit cell vectors. 
As a general comment concerning the geometry of the vacancy, we obtain results similar to those reported in the literature; 
the structure remains planar, two dangling $\sigma$ orbitals rebond in a new weak bond of whose length depends on the super cell size, ranging from 2.05  \rm{\AA} in the $6 \times 6$ case to  1.93 \rm{\AA}  in the $15 \times 15$ one.  The third $\sigma$ orbital remains non-bonded.  In Fig. \ref {Atomicstructure}
we show a reconstructed vacancy in the middle of a 5x5 supercell which, in the calculations, is periodically 
repeated in two dimensions. 

\begin{figure}
\includegraphics[width=0.8\linewidth]{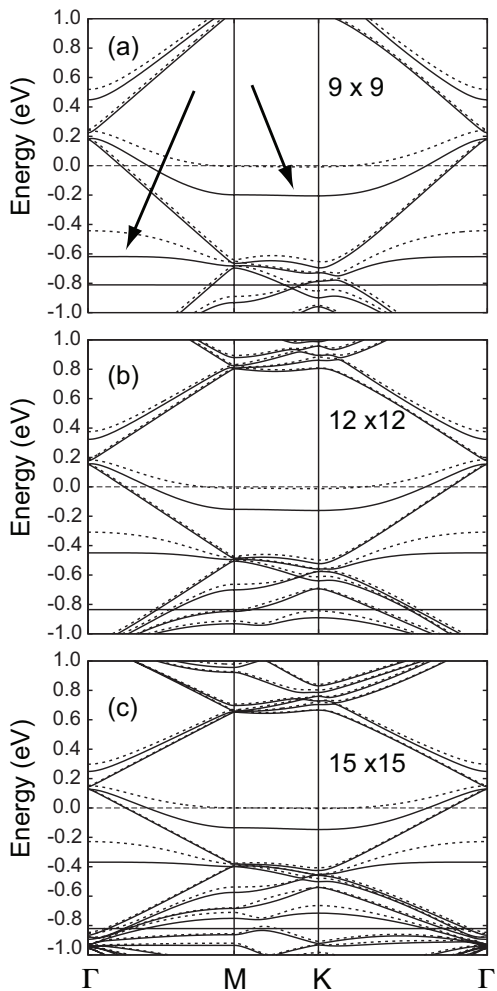}
  \caption{Calculated spin resolved band structure at a vacancy in increasingly $n\times n$ 
supercells in monolayer graphene. Panels (a), (b) and (c) stand for $9 \times 9$, $12\times 12$ and $15\times 15$ supercells 
respectively. Solid and broken lines indicate spin majority and spin minority electronic states. The zero of energy is at 
the Fermi level. The arrows in panel (a) indicate states induced by the defect [similar bands can be identified in panels 
(b) and (c)].}
\label{mono-bands}
\end{figure}

\section{Monolayer graphene}
Although we are interested in the case of an isolated vacancy,   
we are forced to consider a finite concentration of vacancies on the same sublattice; this is what one actually does in 
supercell calculations when using electronic structure codes such as SIESTA. 
As already briefly discussed in the introduction, 
we have first carried out a 6X6 supercell calculation in two cases: i) with H atoms saturating the 
$\sigma$ dangling bonds [Fig. \ref{mono3H}(a)] and ii) 
with no extra H atoms [Fig. \ref{mono3H}(b)]. In the former case, which is not so relevant from an experimental point of view,
the full passivation of the three dangling bonds almost completely prevents the reconstruction of the lattice while in the
latter a strong Jahn-Teller distortion takes place (see \ref {Atomicstructure}).
Left panels show the band structure and right panels the total density of states (DOS).
Only $\pi$ bands are visible in Fig. \ref{mono3H}(a) whereas the unsaturated $\sigma$ dangling bond forms a
band at $\approx$ -0.75 eV in Fig. \ref{mono3H}(b). This band 
presents a large spin splitting and is almost flat as corresponds to a highly localized state. 
On the other hand the splitting of the $\pi$ band is much smaller in both cases 
and presents a visible dispersion which is due to the always present interaction 
between vacancies due to the semi-localized character of the $\pi$ state created by the vacancy\cite{pereira:036801}.
In the DOS the majority spin $\sigma$ peak and the two spin-resolved peaks coming from the $\pi$ state are visible. 

For the fully saturated vacancy neither the unoccupied band nor the occupied one crosses the Fermi level.
The magnetic moment is thus always quantized to 1 $\mu_B$ (we have checked that this is case for 
any concentration of vacancies), as predicted by Lieb's theorem.
Note that the $\sigma$ bonds are saturated and do not host any magnetism here. Note also that 
the bipartite nature of the lattice is preserved, essentially restoring the electron-hole symmetry. The low-energy physics
resulting from these type of vacancies is completely equivalent to the physics of hydrogenated graphene\cite{PhysRevB.81.165409} 
and has been discussed at length in Ref. \onlinecite{palacios:195428}.
For the unsaturated vacancy the band structure presents subtle differences.
Both spin-split $\pi$ bands cross the Fermi level, the upper one actually staying pinned to it. This 
prevents the magnetic moment
from reaching the saturation value of 2$\mu_B$ (1 $\mu_B$ from the $\sigma$ bond plus 1 $\mu_B$ from the vacancy-induced 
$\pi$ state), yielding a total value of around 1.71 $\mu_B$ for this particular calculation. 
This is linked to the remarkable fact that
the Fermi level lies below the Dirac point, which is equivalent to saying that the vacancy acts as an acceptor impurity.
\begin{figure}
\includegraphics[width=1.0\linewidth]{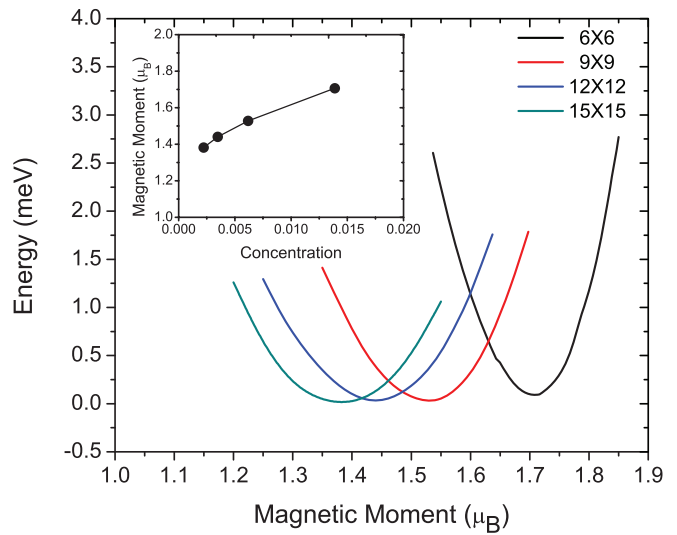}
\caption{(Color online) Total energy versus magnetic moment for different supercell sizes 
for a monovacancy on a graphene monolayer.  The inset indicates the 
magnetic moment at the total energy minimum as a function of the vacancy concentration (inverse of the supercell size).}
\label{susceptibility}
\end{figure}

We should note at this point that the value of the magnetic moment for the reconstructed vacancy,
which is the relevant case from the experimental point of view,  changes with the size of the supercell so
we set out now to do a systematic study.  
Figure \ref{mono-bands} shows the band structure for an increasing supercell size sequence $3n\times 3n$ up to 
$15\times 15$. While the $\sigma$ band becomes quickly completely flat at around -0.8 eV,
the $\pi$ band retains the dispersion and the spin splitting although these become flatter and smaller, respectively,
as the supercell size increases. This reflects the increasing distance between vacancies and the concomitant 
increasing extension of the $\pi$ state induced by the vacancy at the Dirac point. 
(The lattice reconstruction does not allow us to establish a perfect analogy with the Dirac state in the standard 
tight-binding model which decays as $1/r$, but we have no reason to expect otherwise).
Interestingly, the partially occupied upper spin-split $\pi$ band stays pinned at the Fermi level on a part of the Brillouin zone
for all supercell sizes. Also the difference between the Dirac point and the Fermi level decreases which can be easily 
understood since the concentration of vacancies (or acceptor impurities) decreases. It is important at this point to notice that the vacancy does
not only induce one $\pi$ band around the Fermi energy. As indicated by arrows in 
Fig. \ref{mono-bands} (a), a new set of $\pi$ bands emerge which carry spectral weight of the resonance mostly in the occupied part 
of the spectrum. In the zero-concentration limit, this set of bands should merge into a continuum and give a finite a width to
the resonance in the energy sector of occupied states (see below).  

\begin{figure}
\includegraphics[width=0.8\linewidth]{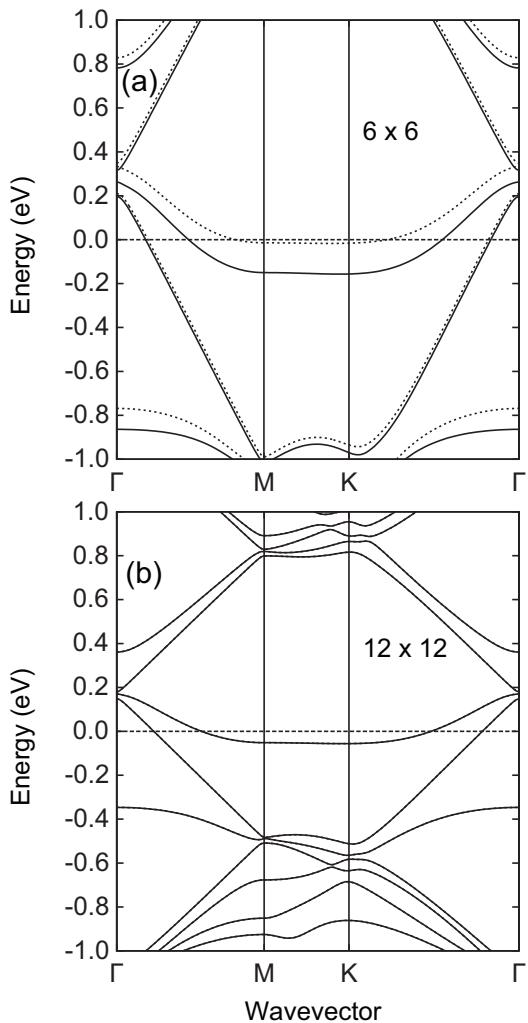}
  \caption{Band structure when the $\sigma$ dangling orbital of the vacancy is saturated with a H atom for a (a) $6\times6$ and  a
(b) $12\times12$ supercell. Solid and broken lines indicate spin majority  and spin minority electronic states. 
The zero of energy is at the Fermi level.  }
\label{mono-sat}
\end{figure}

In Fig. \ref{susceptibility} we show total energy calculations as a function of fixed magnetic moment $\mu$
for various supercell sizes. One can easily appreciate how, as the supercell size increases,
the minimum energy value of $\mu$, $\mu_0$, moves towards 1$\mu_B$, which is the lower limit imposed by the unpaired electron of 
the $\sigma$ dangling bond. 
At the same time, $d^2E(\mu)/d\mu^2|_{\mu_0} \rightarrow 0$, which amounts to a very large susceptibility per vacancy. 
This shallow variation of the energy with the magnetic moment is a remarkable fact; it should be noticed that, for instance,
in the $15 \times 15$ case, the magnetic moment in the $\pi$ states  can vary around 0.4 $\mu_{B}$ within 1 meV. 
This indicates that, even at low temperatures, the magnetic moment is ill defined. This is even more pronounced in the bilayer case (see below).  
The dependence of $\mu_0$ on the inverse of the supercell size (i.e. the concentration of impurities) is plotted in the inset 
of Fig. \ref{susceptibility} (see below for further analysis). 

We finally examine the possibility of having the dangling 
$\sigma$ bond saturated with atomic H. The calculations are performed allowing relaxations of all the atoms as indicated above.  
Now the $\sigma$ band disappears from the energy window of interest (see  Fig. 
\ref{mono-sat}) {\em along with the associated
magnetic moment}.  While for small supercells (or high concentrations of vacancies)
the spin splitting of the $\pi$ band is still visible, it already completely vanishes
for supercell sizes as those considered in the previous case. This result reflects the 
importance of considering the mutual influence between the $\sigma$ and $\pi$ electrons, 
at least as far as magnetic properties is concerned,  when the former are not part
of a bond to other species such as, e.g., H. This effect cannot be 
captured by the Hubbard model where the saturation of the sigma bonds is always implied 
even if the hopping terms are adapted to the atomic reconstruction\cite{PhysRevB.79.075413}.

\section{Bilayer graphene}
\begin{figure}
\includegraphics[width=0.8\linewidth]{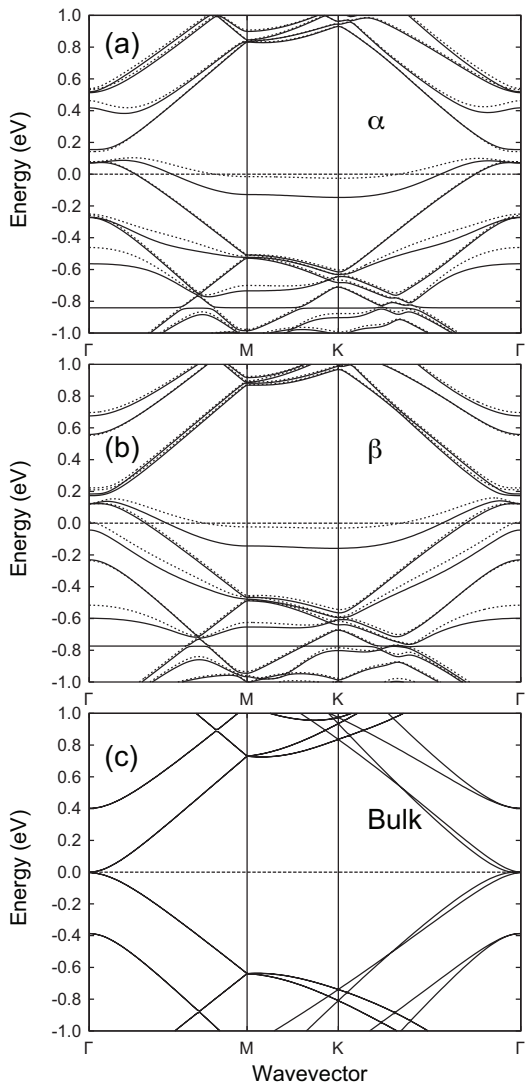}
  \caption{Band structure for vacancies on a bilayer for a $9 \times 9$ supercell.
Solid and broken lines indicate spin majority and spin minority
 electronic states. The zero of energy is at the Fermi level. Panels (a) and (b) refer to a vacancy created on top ($\alpha$) or
hollow ($\beta$) positions, respectively. Panel (c) represents the defect free graphene bilayer.}
\label{bilayer}
\end{figure}

\begin{figure}
\includegraphics[width=1.0\linewidth]{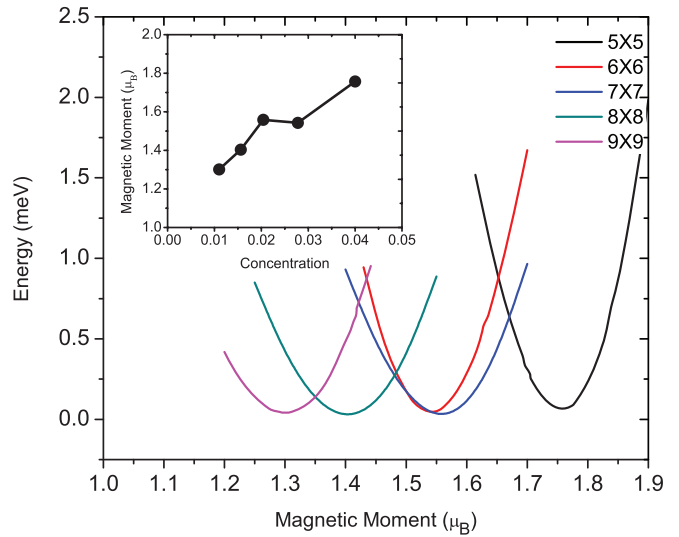}
  \caption{(Color online)  Total energy versus magnetic moment for different supercell sizes for an $\alpha$ vacancy in a graphene bilayer. 
The inset shows the magnetic moment at the total energy minimum as a function of the vacancy concentration (inverse of the supercell size).}
\label{bimom}
\end{figure}

We  now consider vacancies on bilayer graphene with Bernal stacking. In this situation, removing a C atom from one sublattice
or the other is different due to the underlying graphene layer,  resulting in two types of vacancies, $\alpha$ and $\beta$, 
depending on whether or not the vacancy is created on top or hollow position \cite{PhysRevLett.104.096804}. 
The band structure of a vacancy in a $9\times9$ supercell in both cases is shown in Fig.  \ref{bilayer}. 
As in the previous case we allow for full relaxation of the atomic coordinates on the layer containing the vacancy. 
Figure \ref{bilayer} (c) shows the bands for a bilayer without vacancies. The mass acquired by the Dirac electrons 
(the parabolic dispersion at the Fermi energy)
 as a result of the interlayer interactions is evident in the plot. The $\pi$ 
bands associated with the vacancy, regardless of the sublattice creation site, 
are spin-split for small cells. Contrary to the monolayer case, the minority spin band crosses the Fermi level, 
already indicating a stronger tendency towards the quenching of $\pi$ magnetism than in the monolayer.
Therefore, it should not come as a surprise that, as in the monolayer case, the spin splitting goes
to zero as the distance between vacancies increases, remaining only the magnetic moment associated with the $\sigma$ bond. 
The difference between the $\alpha$ and $\beta$ cases is minor. The bands in the $\alpha$ case are narrower than 
those of the $\beta$ case as expected \cite{PhysRevLett.104.096804}.  
The inset in Fig. \ref{bimom} shows $\mu_0$ as a function of the inverse of the supercell size for $\alpha$ vacancies.  
Compared to the monolayer result in Fig. \ref{susceptibility}, one can safely extrapolate  $\mu_0 \rightarrow
1\mu_B$ in the zero-concentration limit. Note that the somewhat erratic behavior of $\mu_0$ as 
a function of the inverse supercell size can be attributed to considering all consecutive sizes while in the monolayer case
we are only plotting results for $3n\times3n$ supercells.
In the light of the results, as for the monolayer, we also expect $\mu_0 \rightarrow 0 \mu_B$ if the $\sigma$ 
dangling bond is saturated. Also, as in the monolayer case, we obtain shallow energy curves versus magnetic moment,
indicating an even higher susceptibility (see Fig. \ref{bimom}).

\section{Critical analysis of results and final considerations}
\begin{figure}
\includegraphics[width=1.0\linewidth]{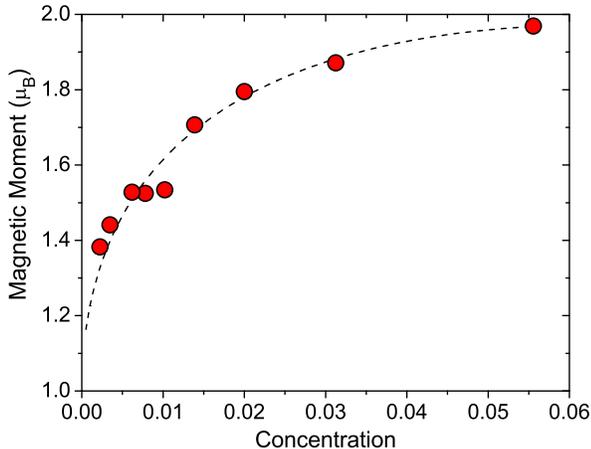}
  \caption{(Color on line). Calculated magnetic moment induced by a vacancy in a graphene monolayer for various concentrations 
(inverse of the supercell size). The red circles are the calculated values and the broken line is a fit to $1+an+b\sqrt{n}$.}
\label{Fit}
\end{figure}

While the results for the bilayer vacancy and the monolayer vacancy with H
seem conclusive regarding the vanishing value of the $\pi$ magnetism in the low 
concentration limit, the ones
for the H free monolayer vacancy remain less clear. We would like to discard any possible influence on the results 
of the specific sequence of supercells considered in our calculations so
we have also performed additional calculations with supercells out of the main sequence $3n\times 3n$. 
We now plot in Fig. \ref{Fit} all the results for the total magnetic moment, including values 
obtained with all types of supercells.
In the light of this plot we can safely conclude that $\mu_0$ goes to 1 $\mu_B$ in the zero-concentration limit $n\rightarrow 0$,
posibly as  $\propto n^\delta$ where $\delta < 1$. In fact, despite that the 
behavior of $\mu_0$ is not monotonic, a good fit to $\mu_0(n)=1+an+b\sqrt{n}$ can be done ($a=-15.11, b=7.64$).

We would like to address now the influence of the 
periodicity on the results. To this aim, we have also performed calculations for truly isolated vacancies 
with the help of the ALACANT package\cite{ALACANT}; in particular,
we have employed our code ANT.G which interfaces with GAUSSIAN09\cite{g09}.
In this case the supercell is sorrounded by an effective medium defined by a two-dimensional 
Bethe lattice\cite{Cluster-Bethe} of coordination three and Slater-Koster parameters for the C $sp$ orbitals . 
Here the Green's function of the supercell is selfconsistently computed subject to a fixed selfenergy representing
the Bethe lattice. In contrast to the calculations with SIESTA, the vacancy is here trully isolated, but the electronic
structure outside the cell remains fixed and unmagnetized.
Unlike bulk graphene, the Bethe lattice model presents a finite density of states at the Fermi energy, which gives
the quasi-localized $\pi$ state of the vacancy a finite lifetime for any cell size even at zero energy.
We have also used here the generalized gradient approximation through the BPBE functional as implemented in 
GAUSSIAN09\cite{g09} and a basis set equivalent to that in the SIESTA calculations. The atomic structure has also been optimized, 
obtaining essentially the same geometry. 
The values of the magnetic moments so obtained are all in the range $\approx 1.1 - 1.3 \mu_B$, with a clear 
trend towards $1.0 \mu_B$ as the system size increases. One may conclude that the periodicity, if anything, enhances the values
of the $\pi$ magnetic moments induced by the vacancy.

\begin{figure}
\includegraphics[width=0.8\linewidth]{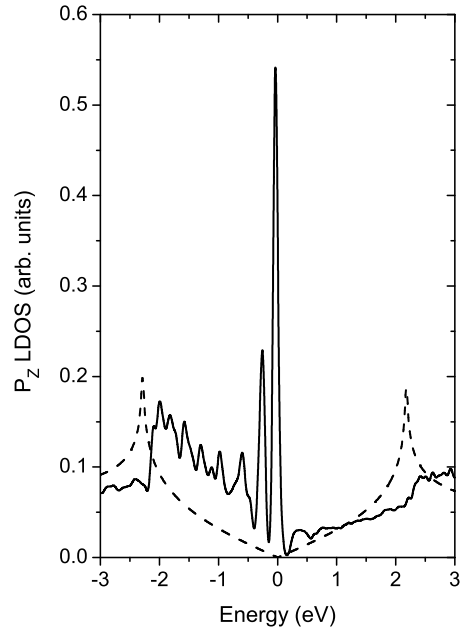}
  \caption{Density of states projected on the $\pi$ orbital at an atom in the vacancy (solid line) in the non-magnetic solution. 
A small (0.05 eV) gaussian broadening has been included for presentation purposes. The corresponding density of states in 
a defect free graphene is represented by the broken line.}
\label{LDOS18X18}
\end{figure}

To make connection with available experimental information\cite{PhysRevLett.104.096804}, we have plotted in Fig. \ref{LDOS18X18} 
the DOS projected on the $\pi$ orbital of the vacancy atom with the dangling bond (for the other two vacancy atoms the results 
are similar). The calculation is a non spin resolved one for a $18 \times 18$ supercell. We obtain an asymmetric 
and almost fully occupied sharp-peaked resonance at the Fermi level, its spectral shape strongly deviating 
from a symmetric Breit-Wigner or $1/\left | E \right |$  resonance\cite{pereira:036801}. Most of its 
weight is in the valence band with no extra structure in the 
conduction band and a small gap right above the main peak. This anomalous form of the line shape is a dramatic 
consequence of the electron-hole symmetry breaking (see qualitatively similar results in a model calculation by 
Pereira et al.\cite{pereira:036801}). The asymmetry and the presence of the small gap right above the sharp peak would prevent, 
in the isolated vacancy limit, the Stoner instability and the formation of an extended magnetic moment.
 
From these results several conclusions can be extracted regarding various experimental observations:

\begin{itemize} 
\item Our results indicate that only a high concentration of ordered vacancies {\em on the same sublattice} can sustain
finite values of the $\pi$ magnetic moments and lead to a ferromagnetically ordered state. The concentration below which
these magnetic moments disappear depends on whether or not the $\sigma$ dangling bond is passivated, being much higher for the 
passivated case. In addition, one should not forget that, in average, the same number of vacancies are expected on both 
sublattices. In this case the $\pi$ magnetic moments are quenched  when vacancies are in proximity\cite{palacios:195428}, further
disfavoring the existence of these magnetic moments. On top of that, an excessive concentration of vacancies will likely render
graphene unstable.  

\item Although one should keep in mind that the STS results by Brihuega et al. \cite{PhysRevLett.104.096804}
refer to surface graphite, our results are compatible with their observations without invoking the existence of $\pi$ magnetism. 
In their experiment no trace of two spin-split peaks near the Fermi energy can be seen. Furthermore, although
the DOS in Fig.  \ref{LDOS18X18} corresponds to a graphene monolayer, the asymmetry in 
the experimental $dI/dV$ peak at low bias nicely compares with our result.
We should note, nevertheless, that we obtain a large magnetic susceptibility mainly associated to the 
soft position of the spin-majority peak in the DOS. The possibility for
thermal fluctuations to wash out this peak from the DOS, masking the spin-split structure cannot be entirely ruled out.

\item As shown in Fig. \ref{mono-bands}, the $\sigma$ band becomes rapidly flat as the concentration of 
vacancies decreases. This indicates
that these localized $\sigma$ magnetic moments do not interact for any reasonable concentration and should behave as paramagnetic 
centers. Upon completion of this work an experiment by Nair et al.\cite{2011arXiv1111.3775N} has unambiguously shown
the paramagnetic behavior of irradiated graphene, ruling out any possibility of magnetic order induced by vacancies and in
complete agreement with our results.

\item To conclude, one should keep in mind that $\pi$ magnetism and magnetic order can still emerge through atomic H adsorption 
or through any other adsorbate capable of similar covalent bonding to $p_z$ orbitals. This magnetism 
should be amenable to experimental verification, for instance in magnetotransport measurements, as recently 
proposed\cite{PhysRevLett.107.016602,nn200558d}.
\end{itemize}

\acknowledgements
We are indebted to I. Brihuega, J. M. G\'omez-Rodr\' iguez and M. M. Ugeda for discussions and helpful comments 
concerning their experimental results. We also appreciate discussions with G. G\'omez-Santos and J. M. Soler.
This work has been financially supported by MICINN of Spain under Grants Nos. FIS2009-12712, FIS2010-21883, MAT07-67845, 
CONSOLIDER CSD2007-00010, and CONSOLIDER  CSD2007-00050.

%

\end{document}